\documentstyle[aps
,epsfig%
,preprint%
]{revtex}
\newcommand{\nn}{\nonumber \\}

\newcommand{\Bx}{\bbox{x}}

\newcommand{\Bg}{\bbox{g}}

\newcommand{\Bp}{\bbox{p}}

\newcommand{\Bu}{\bbox{u}}
\newcommand{\Bv}{\bbox{v}}
\newcommand{\Br}{\bbox{r}}

\newcommand{\Neq}{\mbox{$\,/\!\!\!\!\!=\,$}}
\newcommand{\Tr}{\mbox{Tr}}

\newcommand{\CV}{{\cal V}}
\newcommand{\CG}{{\cal G}}
\newcommand{\BCG}{\bbox{\cal G}}
\begin{document}
\draft
\preprint{Guchi-TP-008}
\date{\today
}
\title{
Equation of state for a classical gas of BPS black holes
}
\author{Nahomi~Kan${}^{1}$%
\thanks{Email address: {\tt b1834@sty.cc.yamaguchi-u.ac.jp}},
Takuya~Maki${}^{2}$%
\thanks{Email address: {\tt maki@clas.kitasato-u.ac.jp}}
and
Kiyoshi~Shiraishi${}^{1,3}$%
\thanks{Email address: {\tt shiraish@sci.yamaguchi-u.ac.jp}}
}
\address{${}^{1}$Graduate School of Science and Engineering, 
Yamaguchi University, 
Yoshida, Yamaguchi-shi, Yamaguchi 753-8512, Japan \\
${}^{2}$Japan Woman's College of Physical Education, 
Kita-karasuyama, Setagaya, Tokyo157, Japan \\
${}^{3}$Faculty of Science, Yamaguchi University,
Yoshida, Yamaguchi-shi, Yamaguchi 753-8512, Japan
}
\maketitle
\begin{abstract}
A point particle treatment to the statistical mechanics of 
BPS black holes in Einstein-Maxwell-dilaton theory is developed.
Because of the absence of the static potential, 
the canonical partition function for $N$ BPS black holes
can be expressed by the volume of the moduli space for them.
We estimate the equation of state for a classical gas of BPS black holes
by Pad\'e approximation and find that the result agrees with the one obtained
by the mean-field approximation. 
\end{abstract}



\section{Introduction}

The thermodynamics of self-gravitating system has some importance
in the context of cosmology.
The long-range nature of gravity gives rise to the breakdown of 
the conventional description of statistical properties, such as
the additivity of the free energy or other `extensive' thermodynamical
quantities.
While the statistical treatment of non-equilibrium systems has recently 
been developed, the existence of the long range force seems essential 
for explaining the formation of the peculiar local structure in the system.

Recently, de Vega and S\'anchez studied the statistical mechanics of the
self-gravitating particle gas\cite{VS}.
They showed that the `thermodynamic limit' must be taken as $N\rightarrow\infty$
with $N/\CV^{1/3}$ fixed in three dimensions, where $N$ is the number of particles
and $\CV$ is the volume of the system. This treatment is required by the
long-range nature of gravity, because there is no {\it ad hoc} cut-off scale.
\footnote{Actually, the horizon length is a candidate for the cut-off scale in the
universe.}

On the other hand, two of the present authors have studied the statistical
mechanics of the well-separated charged particles in Einstein-Maxwell-scalar
theory\cite{SM}. The system was found to be unstable in general, though
there exists a critical case in which the static forces are cancelled each other.

In general relativity, the critical case was investigated and
exact static solutions which describe multi-black hole systems have been obtained
\cite{Papa,Madj,Myer,Shir}. In such a case, the attractive forces (the
gravitational
and scalar-mediated forces) and the repulsive forces (the electrostatic force)
between black holes are
exactly cancelled in the static limit.
In this case, the energy of the system was calculated for small velocities but
for small distances between the black holes\cite{GiRu,FeEa,TrFe,Shi1,Shi2,DS}.
\footnote{For various specific models and their applications, see 
\cite{GiKa,BRO,GPS,KaMi,Mich,MMS,MiSt,MSS,GP1,GP2,SCQM}.}

In this paper, we investigate the multi-black hole system by adopting the technique
of de Vega and S\'anchez\cite{VS}.
Note that a mass-charge relation is satisfied for each individual black hole.
Such extreme black holes, or so-called BPS black holes
\footnote{In \cite{GKLTT}, it is shown that they
saturate a gravitational analogue of the Bogomol'nyi bound on their mass and
charges.},  often appear in string
theories. Besides being of theoretical (academic, methodological) interest, the
study on the thermodynamical aspects of them may be applied to the scenario of
`string cosmology'.

\section{`BPS black holes' in slow motion}

We consider `BPS black holes' in $(d+1)$ dimensional
Einstein-Maxwell-dilaton theory, which is governed by the action
\begin{eqnarray}
 S &=& \int\!\! d^{d+1}
x\,\frac{\sqrt{-g}}{16\pi G}\:\Bigl[\,R-\frac{4}{d-1}\nabla_{\mu}\phi%
\nabla^{\mu}\phi - e^{-\frac{4a}{d-1}\phi} F_{\mu\nu}F^{\mu\nu}\Bigl]\, .
\end{eqnarray}
Here $R$ is the scalar curvature and
$F_{\mu\nu}=\partial_{\mu}A_{\nu}-\partial_{\nu}A_{\mu}$ 
($\mu, \nu=0,1,\dots,d$), while $\phi$ is a dilaton field and
$a$ is a coupling constant.
$G$ denotes the $(d+1)$ dimensional Newton's constant. We set $c=1$.

This theory admits static multi-centered solutions whose metric and
field configurations are given by\cite{Shir}
\begin{eqnarray}
 ds^2 &=& - V^{-\frac{2(d-2)}{d-2+a^2}}dt^2 + 
V^{\frac{2}{d-2+a^2}} d \Bx^2\, ,\\
 e^{-\frac{4a}{d-1}\phi} &=& V^{\frac{2 a^2}{d-2+a^2}} \, ,\\
 A_0(\Bx) &=& \sqrt{\frac{d-1}{2(d-2+a^2)}}\Bigl(1 -\frac{1}{V}\Bigr) \, ,\\
A_i(\Bx)&=&0 \qquad\qquad (i=1,\dots , d)\, ,
\end{eqnarray} 
in terms of a harmonic function
\begin{equation}
V(\Bx)=1+\frac{2(d-2+a^2)}{(d-1)(d-2)}\frac{4\pi}{A_{d-1}}\,G
\sum_{a}~ \frac{m_a}{|\Br_a|^{d-2}}\, ,
\label{eq:V}
\end{equation}
where $\Br_a=\Bx-\Bx_a$, $\Bx_a$ is the position vector of `point particle' $a$
and the labels of `particles' $a, b, \dots$ run over $1, \dots, N$.
$A_{d-1}\equiv 2\pi^{d/2}/\Gamma(d/2)$ is the volume of a unit $d-1$ sphere.

The static point sources corresponding to the `particles' can be described by
the action
\begin{equation}
I = \,- \sum_a\int\!\! ds_a\:\Bigl[m_a\,e^{\frac{2a}{d-1}\phi}+
e_a A_\mu \frac{dx^\mu}{ds_a}
\Bigr]\, ,
\end{equation} 
with the relation
\begin{equation}
\frac{e_a}{m_a} = \sqrt{\frac{2 (d-2 + a^2)}{d-1}}\, .
\end{equation}
This relation is just the extremity condition for spherically
symmetric black holes in the Einstein-Maxwell-dilaton theory\cite{GiMa,GHS}.
Thus the solution can be viewed as the one for an
`$N$ BPS black hole' system, where 
the electric force is cancelled by the forces mediated by the 
graviton and the dilaton.
\footnote{If a dilaton field is coupled, the extreme limit of the black
hole solution has a singularity in general. Therefore, we use the
`quotation marks' around the BPS black hole in this paper.}

While there is no static force between the `black holes',
a velocity dependent force arises if one gives them small velocities
$\Bv_a$.
Many authors calculated the effective theory to $O(v^2)$ for
various types of `black holes'.
One finds the energy of the `BPS black holes' in the 
Einstein-Maxwell-dilaton theory~\cite{DS}
\begin{equation}
H =\sum_{a}\sum_{b}v^{ak}v^{b\ell}
(\delta^i_{k}\delta^j_{\ell}+\delta_{k\ell}\delta^{ij}-
\delta^j_{k}\delta^i_{\ell})
\partial_{ai}\partial_{bj}L \, ,
\end{equation}
where
\begin{equation}
L =-\frac{1}{32\pi G}\int d^d\Bx~V^{\frac{2(d-1)}{d-2+a^2}}(\Bx) \, ,
\end{equation}
with $V$ given by (\ref{eq:V}) and 
the spatial indices $i, j, \dots$ run over $1, \dots, d$.

\section{The canonical partition function}

The canonical partition function for $N$ identical particles at temperature 
$T$ in $d$ spatial dimensions is
\begin{equation}
Z_N=\frac{1}{N!}\frac{1}{h^{Nd}}\int [{\rm dq}][{\rm dp}] \exp\left(-\beta
H\right)\, ,
\end{equation} 
where
$[{\rm dp}]=\prod_{a=1}^N d^d\Bp_a$ and
$[{\rm dq}]=\prod_{a=1}^N d^d\Bx_a$.
$\beta=1/T$, where $h$ is Planck's constant while
Boltzmann's constant is set to be unity.

Because there is no static forces, the Hamiltonian $H$ for 
the system of $N$ `BPS black holes' can be written as
\begin{equation}
H=\frac{1}{2}\sum_{A}\sum_{B}~v^{A}\CG_{AB}v^{B}
=\frac{1}{2}\Bv^T \BCG\Bv\, ,
\end{equation}
where the label $\{A\}=\{(a i)\}$ denotes the combination of the label of the
particle and the spatial index.

The canonical momentum can be found as
\begin{equation}
p_A=\frac{\partial H}{\partial v^A}=\sum_B\CG_{AB}~v^B=(\BCG\Bv)_A \, ,
\end{equation}
and then the Hamiltonian is reexpressed as
\begin{equation}
H=\frac{1}{2}\sum_{A}\sum_{B}~p_{A}\CG^{AB}p_{B}
=\frac{1}{2}\Bp^T \BCG^{-1}\Bp \, .
\end{equation}

Thus the partition function becomes
\begin{equation}
Z_N=\frac{1}{N!}\left(\frac{2\pi}{\beta h^{2}}\right)^{\frac{Nd}{2}}
\int [{\rm dq}] \sqrt{\det\BCG} \, .
\end{equation} 

If we define
\begin{equation}
L'=-\frac{1}{32\pi G}\int d^d\Bx\left[V^{\frac{2(d-1)}{d-2+a^2}}(\Bx)-1-
\frac{4}{d-2}\frac{4\pi}{A_{d-1}}\,Gm
\sum_{c}~ \frac{1}{|\Br_c|^{d-2}}\right] \, ,
\end{equation}
with
\begin{equation}
V(\Bx)=1+\frac{2(d-2+a^2)}{(d-1)(d-2)}\frac{4\pi}{A_{d-1}}\,G m
\sum_{a}~ \frac{1}{|\Br_a|^{d-2}}\, ,
\end{equation}
the Hamiltonian for $N$ `BPS black holes' with a common mass $m$
can be rewritten as
\begin{equation}
H=\frac{1}{2}m\sum_{a}\Bv_a^2+\sum_{a}\sum_{b}v^{ak}v^{b\ell}
(\delta_{k\ell}\delta^{ij}+\delta^i_{k}\delta^j_{\ell}-
\delta^j_{k}\delta^i_{\ell})
\partial_{ai}\partial_{bj}L' \, .
\end{equation}

Then we may use
\begin{equation}
\CG_{(ak)(b\ell)}=m\left[\delta_{ab}\delta_{k\ell}+\frac{2}{m}
(\delta_{k\ell}\delta^{ij}+\delta^i_{k}\delta^j_{\ell}-
\delta^j_{k}\delta^i_{\ell})
\partial_{ai}\partial_{bj}L'\right]\equiv m~g_{(ak)(b\ell)} \, ,
\end{equation}
and obtain the expression for the partition function
\begin{equation}
Z_N=\frac{1}{N!}\left(\frac{2\pi m}{\beta h^{2}}\right)^{\frac{Nd}{2}}
\int [{\rm dq}] \sqrt{\det\Bg}\equiv\frac{1}{N!}\left(\frac{2\pi m}{\beta
h^{2}}\right)^{\frac{Nd}{2}}Q\, ,
\end{equation} 
which is proportial to the volume of the moduli space $Q$.
This feature is the same as that of point-particle systems in two
(spatial) dimensions\cite{Manton,ShMa}, in which there is no static force either.

In the present system, the free energy $F$ is expressed as
\begin{equation}
F=-T\ln Z=NT\ln N-NT-\frac{d}{2}NT\ln\frac{2\pi m}{\beta h^{2}}-NT\ln\CV-
T\ln\frac{Q}{\CV^N}\, .
\end{equation}
Therefore the internal energy $E$ is obtained as
\begin{equation}
\frac{E}{N}=\frac{1}{N}\frac{\partial(\beta F)}{\partial\beta}=\frac{d}{2}T\, ,
\end{equation}
which is the same as that of an ideal gas.

\section{The perturbative approach}

To obtain the equation of state, we have to evaluate the volume of the moduli
space $Q$.

First, we try to obtain the volume of the moduli space
perturbatively.
One can rewrite this as
\begin{eqnarray}
Q&\equiv&\int [{\rm dq}] \sqrt{\det\Bg}
=\int [{\rm dq}] \exp\left(\frac{1}{2}\Tr\ln\Bg\right) \nn
&=&\int [{\rm dq}]\exp\left[\frac{1}{2}\Tr(\Bu-\frac{1}{2}\Bu^2+\cdots)\right] \nn
&=&\int [{\rm dq}]\left[1+\frac{1}{2}\Tr\Bu-\frac{1}{4}\Tr\Bu^2+
\frac{1}{8}(\Tr\Bu)^2+\cdots\right]\, ,
\end{eqnarray} 
where
\begin{equation}
u_{(ak)(b\ell)}=\frac{2}{m}
(\delta_{k\ell}\delta^{ij}+\delta^i_{k}\delta^j_{\ell}-
\delta^j_{k}\delta^i_{\ell})
\partial_{ai}\partial_{bj}L' \, .
\end{equation}

The term including the trace takes the form:
\begin{equation}
\Tr\Bu=\frac{2d}{m}\sum_a\partial_{ai}\partial_{ai}L' \, ,
\end{equation}
and
\begin{equation}
\Tr\Bu^2=\frac{4}{m^2}
\left[d~\delta^{ij}\delta^{i'j'}+2(\delta^{ii'}\delta^{jj'}-
\delta^{ij'}\delta^{ji'})\right]\sum_a\sum_b
\partial_{ai}\partial_{bj}L'\partial_{ai'}\partial_{bj'}L' \, .
\end{equation}

To obtain the leading and next-to-leading contributions for a small $m$,
we expand $L'$ as
\begin{eqnarray}
L'=-\frac{1}{32\pi}\int d^d\Bx& &\left[\frac{1}{2}
\left(\frac{4}{d-2}\frac{4\pi}{A_{d-1}}\right)^2
\frac{d-a^2}{2(d-1)}Gm^2
\sum_{c}\sum_{d}\frac{1}{|\Br_c|^{d-2}}\frac{1}{|\Br_d|^{d-2}}
\right.\nn
& &\left.+\frac{1}{6}
\left(\frac{4}{d-2}\frac{4\pi}{A_{d-1}}\right)^3
\frac{d-a^2}{2(d-1)}\frac{1-a^2}{d-1}G^2m^3
\sum_{c}\sum_{d}\sum_{e}\frac{1}{|\Br_c|^{d-2}}\frac{1}{|\Br_d|^{d-2}}
\frac{1}{|\Br_e|^{d-2}}\right]\nn
& &+O(m^4)\, .
\end{eqnarray}
Using this, the traces can be written as
\footnote{See Appendix for detailed calculations.}
\begin{eqnarray}
\Tr\,\Bu&=&d\left(\frac{4}{d-2}\frac{4\pi}{A_{d-1}}\right)
\frac{d-a^2}{2(d-1)}Gm
\mathop{\sum\sum}_{a\Neq b}\frac{1}{|\Bx_a-\Bx_b|^{d-2}}\nn
& &+\frac{d}{2}
\left(\frac{4}{d-2}\frac{4\pi}{A_{d-1}}\right)^2
\frac{d-a^2}{2(d-1)}\frac{1-a^2}{d-1}G^2m^2 \nn
& &~\times\left[
\mathop{\sum\sum\sum}_{b\Neq a, c\Neq
a}\frac{1}{|\Bx_a-\Bx_b|^{d-2}}\frac{1}{|\Bx_a-\Bx_c|^{d-2}}+
\mathop{\sum\sum}_{a\Neq b}\frac{1}{|\Bx_a-\Bx_b|^{2(d-2)}}\right] \nn
& &+O((Gm)^3)\, ,
\end{eqnarray}
\begin{eqnarray}
\Tr\,\Bu^2&=&d\left(\frac{4}{d-2}\frac{4\pi}{A_{d-1}}\right)^2
\left(\frac{d-a^2}{2(d-1)}\right)^2G^2m^2 \nn
& &~\times\left[
\mathop{\sum\sum\sum}_{b\Neq a, c\Neq
a}\frac{1}{|\Bx_a-\Bx_b|^{d-2}}\frac{1}{|\Bx_a-\Bx_c|^{d-2}}+
\mathop{\sum\sum}_{a\Neq b}\frac{1}{|\Bx_a-\Bx_b|^{2(d-2)}}\right] \nn
& &+O((Gm)^3)\, .
\end{eqnarray}

Then we can obtain the folowing expression for $Q$:
\begin{eqnarray}
\frac{Q}{\CV^N}&=&
1+\frac{d}{2}\left(\frac{4}{d-2}\frac{4\pi}{A_{d-1}}\right)
\frac{d-a^2}{2(d-1)}\,Gm\, N(N-1)
\frac{1}{\CV^2}\int\!\!\!\int\frac{d^d\Bx_1d^d\Bx_2}{|\Bx_1-\Bx_2|^{d-2}}\nn
& &-\frac{d}{4}
\left(\frac{4}{d-2}\frac{4\pi}{A_{d-1}}\right)^2
\frac{d-a^2}{2(d-1)}\frac{d-2+a^2}{2(d-1)}\,G^2m^2 \nn
& &~\times\left[N(N-1)(N-2)\frac{1}{\CV^3}\int\!\!\!\int\!\!\!\int
\frac{d^d\Bx_1d^d\Bx_2d^d\Bx_3}{|\Bx_1-\Bx_2|^{d-2}|\Bx_1-\Bx_3|^{d-2}}
\right. \nn
& &\left.\qquad+2N(N-1)\frac{1}{\CV^2}\int\!\!\!\int
\frac{d^d\Bx_1d^d\Bx_2}{|\Bx_1-\Bx_2|^{2(d-2)}}\right]\nn 
& &+\frac{d^2}{8}
\left(\frac{4}{d-2}\frac{4\pi}{A_{d-1}}\right)^2
\left(\frac{d-a^2}{2(d-1)}\right)^2 G^2m^2 \nn
& &~\times\left[N(N-1)(N-2)(N-3)\left(\frac{1}{\CV^2}\int\!\!\!\int
\frac{d^d\Bx_1d^d\Bx_2}{|\Bx_1-\Bx_2|^{d-2}}\right)^2
\right. \nn
& &\qquad+4N(N-1)(N-2)\frac{1}{\CV^3}\int\!\!\!\int\!\!\!\int
\frac{d^d\Bx_1d^d\Bx_2d^d\Bx_3}{|\Bx_1-\Bx_2|^{d-2}|\Bx_1-\Bx_3|^{d-2}} \nn
& &\left.\qquad+2N(N-1)\frac{1}{\CV^2}\int\!\!\!\int
\frac{d^d\Bx_1d^d\Bx_2}{|\Bx_1-\Bx_2|^{2(d-2)}}\right]\nn 
& &+O((Gm)^3)\, ,
\end{eqnarray}
where $\CV=\int d^d\Bx$.

This expression includes divergent integrals and thus we must use
the small cut-off length. These terms are, however, eliminated by $1/N$ factors
in the $N\rightarrow\infty$ limit\cite{VS}. 

In the thermodynamical limit, or in the limit of large $N$, we obtain
\begin{eqnarray}
\lim_{N\rightarrow\infty}\frac{1}{N}\ln\frac{Q}{\CV^N}&=&
\frac{d}{2}\left(\frac{4}{d-2}\frac{4\pi}{A_{d-1}}
\frac{GmN}{\ell^{d-2}}\right)\frac{d-a^2}{2(d-1)} 
\frac{\ell^{d-2}}{\CV^2}\int\!\!\!\int
\frac{d^d\Bx_1d^d\Bx_2}{|\Bx_1-\Bx_2|^{d-2}}\nn
& &-\frac{d^2}{2}
\left(\frac{4}{d-2}\frac{4\pi}{A_{d-1}}\frac{GmN}{\ell^{d-2}}\right)^2
\left(\frac{d-a^2}{2(d-1)}\right)^2
\left(\frac{\ell^{d-2}}{\CV^2}\int\!\!\!\int
\frac{d^d\Bx_1d^d\Bx_2}{|\Bx_1-\Bx_2|^{d-2}}\right)^2 \nn
& &+\frac{d^2}{2}
\left(\frac{4}{d-2}\frac{4\pi}{A_{d-1}}\frac{GmN}{\ell^{d-2}}\right)^2
\left(\frac{d-a^2}{2(d-1)}\right)^2
\left(1-\frac{d-2+a^2}{2d(d-a^2)}\right) \nn
& &\quad\times\frac{\ell^{2(d-2)}}{\CV^3}\int\!\!\!\int\!\!\!\int
\frac{d^d\Bx_1d^d\Bx_2d^d\Bx_3}{|\Bx_1-\Bx_2|^{d-2}|\Bx_1-\Bx_3|^{d-2}} \nn
& &+O\left(\left(\frac{GmN}{\ell^{d-2}}\right)^3\right)\, ,
\label{eq:30}
\end{eqnarray}
where we have introduced a length scale $\ell$ and the value $N/\ell^{d-2}$ is
fixed.

For a spherical box of radius $\ell$, on can find
\begin{equation}
\frac{\ell^{d-2}}{\CV^2}\int\!\!\!\int
\frac{d^d\Bx_1d^d\Bx_2}{|\Bx_1-\Bx_2|^{d-2}}=\frac{2d}{d+2}\, ,
\label{eq:31}
\end{equation}
\begin{equation}
\frac{\ell^{2(d-2)}}{\CV^3}\int\!\!\!\int\!\!\!\int
\frac{d^d\Bx_1d^d\Bx_2d^d\Bx_3}{|\Bx_1-\Bx_2|^{d-2}|\Bx_1-\Bx_3|^{d-2}}
=\frac{d(5d+2)}{(d+2)(d+4)}\, ,
\label{eq:32}
\end{equation}
where $\CV=A_{d-1}\ell^d/d$.

The pressure $P$ is derived from the free energy $F$ as
\begin{equation}
P=-\frac{\partial F}{\partial\CV}\, ,
\end{equation}
thus we find
\begin{equation}
\frac{P\CV}{NT}=1+\frac{\CV}{N}\frac{\partial}{\partial\CV}
\ln\frac{Q}{\CV^N}
=1+\frac{\ell}{Nd}\frac{\partial}{\partial\ell}
\ln\frac{Q}{\CV^N}\, .
\label{eq:34}
\end{equation}
Substituting (\ref{eq:30}), (\ref{eq:31}), (\ref{eq:32}) into (\ref{eq:34}), we
obtain the equation of state in the limit of large $N$:
\begin{equation}
\frac{P\CV}{NT}=f(y)\, ,
\end{equation}
with
\begin{equation}
f(y)=1-b_1\, y+b_2\, y^2+O(y^3)\, ,
\end{equation}
where
\begin{equation}
b_1=
\frac{d-2}{2}\left[\frac{2(d-a^2)}{d-1}\right]
\frac{2d}{d+2}\, ,
\end{equation}
\begin{equation}
b_2=d(d-2)
\left[\frac{2(d-a^2)}{d-1}\right]^2\left[
\left(\frac{2d}{d+2}\right)^2
-\left(1-\frac{d-2+a^2}{2d(d-a^2)}\right)
\frac{d(5d+2)}{(d+2)(d+4)}\right]\, ,
\end{equation}
and
\begin{equation}
y=\frac{1}{d-2}\frac{4\pi}{A_{d-1}}\frac{GmN}{\ell^{d-2}}\, ,
\end{equation}
is a dimensionless parameter. $y$ can be regarded as a normalized 
coupling constant of the interaction.

Now we can rewrite this equation of state in the Van der Waals form
\begin{equation}
P\CV'=NT~~~~~\mbox{with}~~~~\CV'=\CV \left[f(y)\right]^{-1}\, ,
\end{equation}
which exhibits an `anti-excluded' volume effect for a small $y$
if $a^2<d$. There is an effective attractive force  between `BPS 
black holes' for $a^2<d$. The effective force is repulsive for $a^2>d$,
while for $a^2=d$ there is no interaction among the `BPS 
black holes'.

\section{The strong coupling limit and Pad\'e approximation for
the equation of state}

In this section, we consider the strong coupling limit, $y\gg 1$.
By counting the maximal number of $G$ appearing in $Q$, we find
\begin{equation}
\frac{Q}{\CV^N}\approx 
y^{\frac{Nd}{2}\frac{d-a^2}{d-2+a^2}}\times const.\, ,
\end{equation}
since $\BCG$ is an $(Nd)\times(Nd)$ matrix.
\footnote{Although the rank of $\BCG$ may be less than $Nd$, this is 
a reasonable aproximation for $N\rightarrow\infty$.}

This leads to, in the large $y$ limit,
\begin{equation}
\frac{P\CV}{NT}\rightarrow 1-\frac{d-2}{2}\frac{d-a^2}{d-2+a^2}=
\frac{d a^2-(d-2)^2}{2(d-2+a^2)}\, .
\end{equation}
If $a^2<\frac{(d-2)^2}{d}$, the pressure $P$ becomes negative!
This means that the gas will be spontaneously condensed for large
$y$.

This feature can be explained by considering the
moduli space metric for two body system\cite{Shi2}.
In the center-of-mass system, the
moduli space metric in the small distance limit is
\begin{equation}
ds^2\approx \left(\frac{Gm}{r^{d-2}}\right)^{\frac{d-a^2}{d-2+a^2}}
(dr^2+r^2d\Omega^2)\, ,
\end{equation}
where $r$ is the distance between two particles and $\Omega$ represents
the angular variables.
One can use a new radial coordinate near the origin:
\begin{equation}
R=\int dr r^{-\frac{(d-2)(d-a^2)}{2(d-2+a^2)}}\, .
\end{equation}
The range of $R$ is $[0,\infty)$ for $a^2\ge\frac{(d-2)^2}{d}$, while
$(-\infty,\infty)$ for $a^2<\frac{(d-2)^2}{d}$.
Therefore in the case of $a^2<\frac{(d-2)^2}{d}$, two particles merges, or
$R\rightarrow -\infty$, for a sufficiently small impact parameter.
This appears to be the cause of the instability of the gaseous state.


An approximation which interpolates the small and large $y$
regions shows
\begin{equation}
\frac{Q}{\CV^N}\approx \left(q_{p1}(y)\right)^N\equiv
(1+\alpha_1\,y)^{\frac{Nd}{2}\frac{d-a^2}{d-2+a^2}}\, ,
\end{equation}
with
\begin{equation}
\alpha_1=\frac{2(d-2+a^2)}{d-1}\frac{2d}{d+2}\, .
\end{equation}

Although this approximation does not utilize the information of
the second order in the perturbative expansion, this form is very useful
and provides the lowest-order Pad\'e approximation for $f(y)$ (see below)
besides it garantees the positivity of the volume of the moduli space
in the simplest way.

In this approximation, the equation of state reads
\begin{equation}
\frac{P\CV}{NT}=f(y)=1-\frac{(d-2)(d-a^2)}{2(d-2+a^2)}
\frac{\alpha_1\, y}{1+\alpha_1\,y}
=1-\frac{(d-2)(d-a^2)}{d-1}\frac{2d}{d+2}
\frac{y}{1+\alpha_1\,y}\, .
\end{equation}

At the critical $y=y_c$, $y_c$ satisfies $f(y_c)=0$, the gas cannot remain
in the normal gaseous phase. In the present approximation scheme, the value of
$y_c$ is
\begin{equation}
y_c=\frac{d+2}{2d}
\frac{d-1}{(d-2)^2-da^2}\, ,
\end{equation}
for $a^2<a_c^2\equiv (d-2)^2/d$.

The next-to-leading Pad\'e approximation for the equation of state 
is also possible. The corresponding form for the moduli space volume is
written by
\begin{equation}
\frac{Q}{\CV^N}\approx \left(q_{p2}(y)\right)^N\equiv
(1+2\alpha_1\,y+\alpha_2\, y^2)^{\frac{Nd}{4}\frac{d-a^2}{d-2+a^2}}\, ,
\end{equation}
with
\begin{equation}
\alpha_2=2\alpha_1^2-\frac{2}{d-2}\frac{d-2+a^2}{d-a^2}b_2\, .
\end{equation}


\section{The mean-field approximation}

In \cite{DS}, the effective field theory of `BPS black holes' was
constructed.
Here we use the effective field theory to obtain the 
equation of state for a gas of `BPS black holes'
by the mean-field approximation.

In the mean-field approximation
the number density with a chemical potential $\mu$ is given by
\begin{equation}
n=\frac{1}{h^d}\int d^dp\, \exp\left[-\beta\left(
\frac{p^2}{2m V^{\frac{d-a^2}{d-2+a^2}}}-\mu\right)\right]
\propto V^{\frac{d(d-a^2)}{2(d-2+a^2)}}\, ,
\end{equation}
where the gauge interaction term is discarded.
\footnote{Non-perturbative effect of the interaction was studied in \cite{DS}.}

The mean-field potential $V$ satisfies the following equation\cite{DS}:
\begin{equation}
\partial^2 V+8\pi\frac{d-2+a^2}{d-1}Gm\, n_0V^{\frac{d(d-a^2)}{2(d-2+a^2)}}=0\, ,
\end{equation}
where we set $n=n_0V^{\frac{d(d-a^2)}{2(d-2+a^2)}}$ with a constant $n_0$.
For the spherical symmetric case, the equation for $V$ reads
\begin{equation}
\frac{d^2 V}{d\tilde{r}^2}+\frac{d-1}{\tilde{r}}\frac{d
V}{d\tilde{r}}+\frac{2d(d-2)(d-2+a^2)}{d-1}\,
y\frac{\CV n_0}{N}V^{\frac{d(d-a^2)}{2(d-2+a^2)}}=0\, ,
\label{eq:53}
\end{equation}
with the boundary conditions
\begin{equation}
\left.\frac{d V}{d\tilde{r}}\right|_{r=0}=0,~~~
\left.\frac{d
V}{d\tilde{r}}\right|_{\tilde{r}=1}=-\frac{2(d-2)(d-2+a^2)}{d-1}y\, ,
\label{eq:54}
\end{equation}
and
\begin{equation}
\left.V\right|_{\tilde{r}=1}=1+\frac{2(d-2+a^2)}{d-1}y\, .
\label{eq:55}
\end{equation}
Here $\tilde{r}=r/\ell$ and we used $\int d^d\Bx\, n=N$ in the
spherical box of which radius is $\ell$.

To obtain the relation between $y$ and $N/(n_0\CV)$, one can use the
scaling covariance. It can be found that one has to solve
\begin{equation}
V_0''(\lambda)+\frac{d-1}{\lambda}
V_0'(\lambda)+\frac{2d(d-2)(d-2+a^2)}{d-1}\,
V_0^{\frac{d(d-a^2)}{2(d-2+a^2)}}(\lambda)=0\, ,
\label{eq:LE}
\end{equation}
with the boundary conditions
\begin{equation}
V_0'(0)=0,\quad V_0(0)=1\, ,
\end{equation}
instead of solving (\ref{eq:53}) with (\ref{eq:54}) and (\ref{eq:55}).
Here the prime (${}'$) denotes the derivative with respect to $\lambda$.

Next, one must solve the equation
\begin{equation}
\lambda_s\frac{V_0'(\lambda_s)}{V_0(\lambda_s)}=
-\frac{(d-2)\frac{2(d-2+a^2)}{d-1}y}{1+\frac{2(d-2+a^2)}{d-1}y}\, ,
\label{eq:ly}
\end{equation}
to get $\lambda_s(y)$.

Finally, $N/(n_0\CV)$ is given by
\begin{equation}
\frac{N}{n_0\CV}= q_{mf}(y)\equiv
-\frac{d-1}{2(d-2)(d-2+a^2)}\frac{V_0'(\lambda_s)}{\lambda_s}
\left(V_0(\lambda_s)+\frac{\lambda_s}{d-2}V_0'(\lambda_s)
\right)^{-\frac{d(d-a^2)}{2(d-2+a^2)}}\, .
\label{eq:qm}
\end{equation}

In other words, solving (\ref{eq:ly}), we have
\begin{equation}
y=
-\frac{d-1}{2(d-2)(d-2+a^2)}\lambda_sV_0'(\lambda_s)
\left(V_0(\lambda_s)+\frac{\lambda_s}{d-2}V_0'(\lambda_s)
\right)^{-1}\, ,
\end{equation}
and this relation with (\ref{eq:qm}) gives an implicit functional expression
in terms of the parameter $\lambda_s$.

Note that for a small $y$, we can find
\begin{equation}
q_{mf}(y)=1+\frac{2d^2(d-a^2)}{(d-1)(d+2)}\, y+O(y^2)\, ,
\end{equation}
which coincides with $q_{p1}(y)$ up to the first order in $y$.

On the other hand, for $y$ grows as
$V_0(\lambda_s)+\frac{\lambda_s}{d-2}V_0'(\lambda_s)\rightarrow 0$,
we find $q_{mf}(y)\propto y^{\frac{d(d-a^2)}{2(d-2+a^2)}}$ in the limit of
large $y$. This behavior agrees with the previous analysis on the
strong coupling region.

In the mean-field approximation, the canonical partition function can
 be expressed as
\begin{equation}
Z_N=\frac{1}{N!}\left[\frac{1}{\CV}\int d^d\Bx\left(\frac{2\pi m}{\beta
h^{2}}V^{\frac{d-a^2}{d-2+a^2}}\right)^{\frac{d}{2}}\right]^N
=\frac{1}{N!}\left(\frac{2\pi m}{\beta
h^{2}}\right)^{\frac{Nd}{2}}\left[q_{mf}(y)\right]^N\,  .
\end{equation}

Then, the equation of state reads $P\CV/(NT)=f(y)$ with
\begin{equation}
f(y)=1-\frac{d-2}{d}y\frac{\partial}{\partial y}
\ln q_{mf}(y)\, .
\end{equation}

We evaluate $f(y)$ by solving Eq.~(\ref{eq:LE}) by a numerical method,
since the solution cannot be given in analytically in general.
\footnote{See, however, for analytic approximations, \cite{Liu,MeRy}.
 See also \cite{NaLB}.}

\begin{figure}
\centering
\mbox{\epsfbox{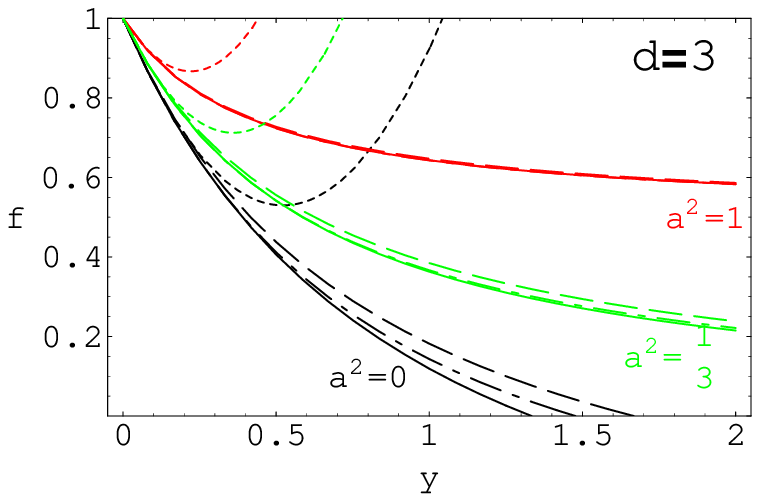}\epsfbox{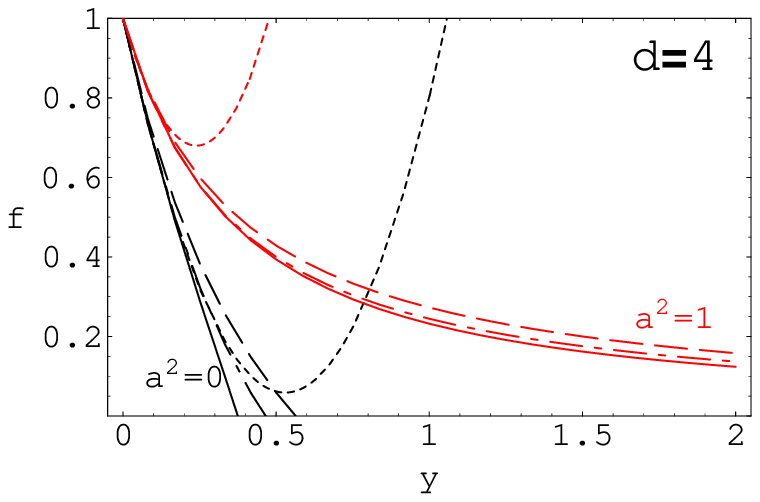}}
\mbox{\epsfbox{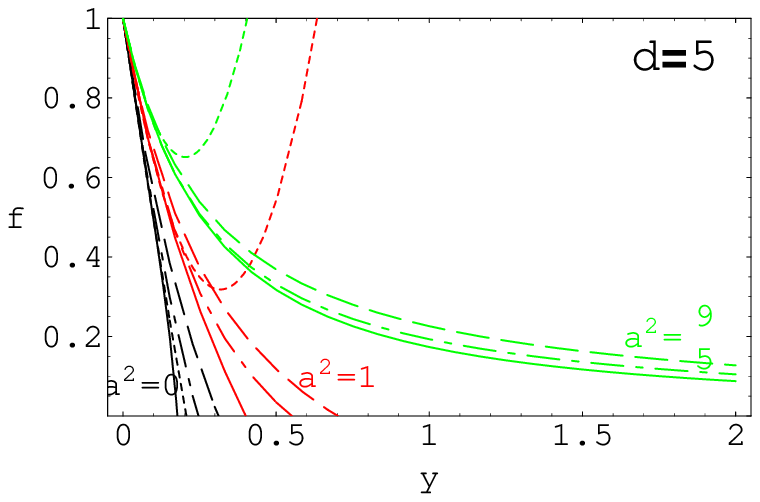}\epsfbox{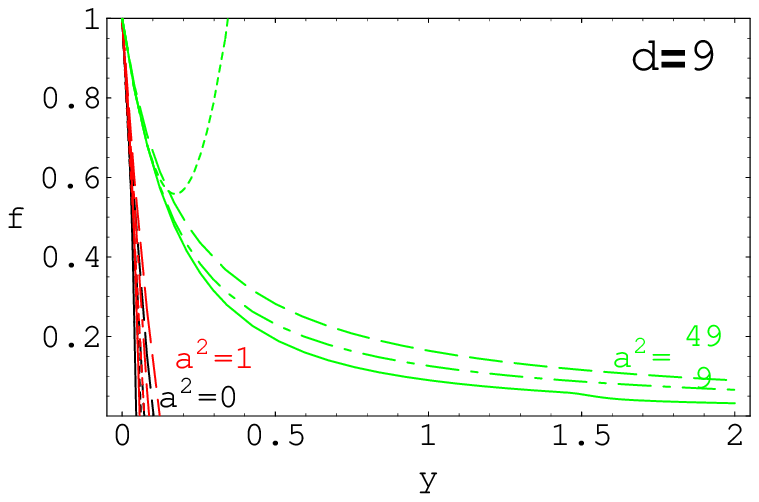}}
\caption{%
$f(y)$ is plotted against $y$ for $d=3$, $d=4$, $d=5$ and $d=9$. 
The solid line is obtained by the mean-field approximation,
the broken line by the lowest-order Pad\'e approximation, the dot-broken line by
the next-to-leading Pad\'e approximation, and the dotted line is the series
expansion up to the second order.}
\label{fig:eos}
\end{figure}

In Fig.~\ref{fig:eos}, $f(y)$ obtained by approximations considered above
is shown. All the mean-field and Pad\'e approximations indicate the consistent
behavior of the function. The difference appears to be small when $d-a^2$ is
small.


There are regions where the pressure decreases with decreasing volume, where the
system cannot be stable and will compress itself  to a smaller volume.
The isothermal compressibility $K$,
\begin{equation}
K=-\frac{1}{\CV}\frac{\partial\CV}{\partial P}=
\frac{\CV}{NT}\left(f(y)+\frac{d-2}{d}y
f'(y)\right)^{-1}\, ,
\end{equation}
becomes negative in such a region.

The specific heat at constant pressure $c_P$ given by
\begin{equation}
c_P=c_{\CV}-\frac{T}{N}
\frac{\left(\frac{\partial P}{\partial T}\right)^2}%
{\frac{\partial P}{\partial\CV}} 
=\frac{d}{2}+
\left(f(y)\right)^2\left(f(y)+\frac{d-2}{d}y
f'(y)\right)^{-1}\, ,
\end{equation}
also becomes negative in such a region.

These two quantities diverges when $y\rightarrow y_0$, and become negative
for $y>y_0$.

The values of $y_c$ ($f(y_c)=0$) and $y_0$ ($K^{-1}=c_P^{-1}=0$) for several
cases are listed in Table~1.
$(mf)$ means that the value is obtained by the mean-field aproximation,
while $(p1)$ and $(p2)$ mean that the values are estimated by the
leading and next-to-leading Pad\'e approximations.

\[
\begin{array}{|l||l|l|l|l|}\hline
  & d=3 & d=4 & d=5 & d=5  \\ 
  & a^2=0 & a^2=0 & a^2=0 & a^2=1  \\ \hline\hline
y_c~(mf)& 1.33354 & 0.374997 & 0.17789  & 0.399981  \\ \hline
y_c~(p2)& 1.47578 & 0.46414 & 0.245029 & 0.549095   \\
\hline y_c~(p1)& 1.66667 & 0.562500 & 0.311111 & 0.700000  \\
\hline\hline y_0~(mf)& 0.944025 & 0.248672 & 0.120393 & 0.235276 \\
\hline y_0~(p2)& 1.04266  & 0.288196 & 0.144323 & 0.292631 \\
\hline y_0~(p1)& 1.17851  & 0.347644 & 0.180799 & 0.374393 \\
\hline
\multicolumn{5}{c}{Table~1}
\end{array}
\]

We have obtained the similar critical values for $y$ by the different
approxmations. Thus the precise values are expected to be in the vicinity
of the approximated values. This will be confirmed by the numerical simulation
in the future.

\section{Particle distributions}

We consider a spherical distribution of the `BPS black holes'.
We define the mass distribution in the mean-field approximation by
\begin{equation}
M(\tilde{r})=A_{d-1}\int_0^{\tilde{r}}\frac{n_0}{N}
[V(x)]^{\frac{d(d-a^2)}{2(d-2+a^2)}} x^{d-1} dx\, .
\end{equation}
$M(\tilde{r})$ gives the fraction of the mass inside the sphere with the radius
$\tilde{r}$.

This can be approximated as
\begin{equation}
M(\tilde{r})\approx\tilde{r}^{D}\, ,
\end{equation}
noting that $M(0)=0$ and $M(1)=1$.

$D$ is considered as the most naive definition of the fractal
dimension. We evaluate the value of
$D$ at the least mean square of the deviations. The results for $d=3, 4, 5$ are
exhibited in Table 2, 3 and 4.

\[
\begin{array}{|l||l|l|l|l|l|l|l|l|l|l|}\hline
\multicolumn{11}{|c|}{d=3}\\ \hline\hline\hline
\qquad y & 0.2 & 0.4 & 0.6 & 0.8 & 1.0 & 1.2 & 1.4 & 1.6 & 1.8 & 2.0 \\
\hline\hline D~(a^2=0)& 2.69 & 2.46 & 2.29 & 2.15 & 2.04 & 1.95  & 1.87 & 1.81 &
1.75 & 1.70 \\
\hline D~(a^2=1/3)& 2.74 & 2.58 & 2.46 & 2.38 & 2.31 & 2.26 &
2.22 & 2.19 & 2.16 & 2.14 \\
\hline D~(a^2=1)& 2.83 & 2.74 & 2.69 & 2.65 & 2.62 & 2.60 &
2.59 & 2.58 & 2.57 & 2.56 \\ \hline
\multicolumn{11}{c}{Table.~2}
\end{array}
\]

\[
\begin{array}{|l||l|l|l|l|l|l|l|l|l|l|}\hline
\multicolumn{11}{|c|}{d=4}\\ \hline\hline\hline
\qquad y & 0.1 & 0.2 & 0.3 & 0.4 & 0.5 & 0.6 & 0.7 & 0.8 & 0.9 & 1.0 \\
\hline\hline D~(a^2=0)& 3.52 & 3.11 & 2.76  & 2.42 & 2.09 & & & &
&\\
\hline D~(a^2=1)& 3.68 & 3.45 & 3.29 & 3.17 & 3.07 & 3.00 &
2.93 & 2.88 & 2.84 & 2.80
\\
\hline
\multicolumn{11}{c}{Table.~3}
\end{array}
\]

\[
\begin{array}{|l||l|l|l|l|l|l|l|l|l|l|}\hline
\multicolumn{11}{|c|}{d=5}\\ \hline\hline\hline
\qquad y & 0.1 & 0.2 & 0.3 & 0.4 & 0.5 & 0.6 & 0.7 & 0.8 & 0.9 & 1.0 \\
\hline\hline D~(a^2=0)& 4.00 & 2.90 &  &  &  & & & &
&\\
\hline D~(a^2=1)& 4.29 & 3.77 & 3.35 & 3.02 & 2.78 &  &  & & &\\
\hline D~(a^2=9/5)& 4.48 & 4.14 & 3.91 & 3.75 & 3.62 & 3.53 & 3.45
  & 3.39 & 3.33 & 3.29 
\\
\hline
\multicolumn{11}{c}{Table.~4}
\end{array}
\]

We find that $D$ slowly decreases as $y$ grows.
These behaviors must be checked by the numerical simulations in the future.

\section{Summary}

In summary,
we have studied many-body system of `BPS black holes' in $d$ dimensions.
The canonical partition function of the system is proportional to the volume of
the moduli space of $N$ `BPS black holes'.
Therefore the difference from an ideal gas arises through the change of the
effective volume of the system: the internal energy and heat capacity at a fixed
volume are the same as those of an ideal gas.
 
We have estimated the equation of state of the gas by perturbations in the
case of a weakly self-interacting gas. The appropriate variable for the expansion
has been found to be
$y\propto Gm/\ell^{d-2}$, where $\ell$ is the length scale of the box containing
the gas.
We have also estimated the equation of state of the gas by Pad\'e approximation
associated with counting the highest order of the couplings, and by the mean-field
method.

In these approximation, we have found that the pressure simply decreases
as the value of $y$ increases if the temperature and the density are fixed.
This feature is common in the cases for the different spatial dimensions and
the dilaton couplings.

For each dimensionality $d$, the critical value for the dilaton coupling $a$ has
been found.
If the absolute value of the coupling is smaller than the critical value
$a_c=\frac{d-2}{\sqrt{d}}$, there exists the critical value $y_c$; for $y>y_c$, the
pressure becomes negative.

A usual thermodynamical limit, $N\rightarrow\infty$, $\CV\rightarrow\infty$ and
$N/\CV$ fixed, can be taken only if $a^2>a_c^2$. This limit can be obtained by
taking
$y\rightarrow\infty$.
The equation of state is then
\begin{equation}
\frac{P\CV}{NT}=\frac{d a^2-(d-2)^2}{2(d-2+a^2)}\, .
\end{equation}
The isothermal compressibility $K$ is
\begin{equation}
K=\frac{\CV}{NT}\frac{2(d-2+a^2)}{d a^2-(d-2)^2}\, ,
\end{equation}
and the specific heat at constant pressure $c_P$ is 
\begin{equation}
c_P=\frac{d}{2}+\frac{d a^2-(d-2)^2}{2(d-2+a^2)}\, .
\end{equation}

Of course the analytical attempt to describe thermodynamic property of the
`BPS black hole' gas may have been inconclusive and numerical simulations
of the self-interacting system may be needed.
The present work provides a guideline for such numerical calculations
to survey the critical points of the `BPS black hole' gas.

For the near-extremal cases, we may treat static potentials as a perturbation from
the critical coupling. These cases will be interesting because they may exhibit
complicated temperature-dependent critical behaviors.



\section*{Acknowledgements}
KS would like to thank Kenji Sakamoto for reading this manuscript.
He also thanks Yoshinori Cho for useful comments.
\newpage

\appendix
\section*{Some useful formulas}

\begin{equation}
\int d^d\Bx~\frac{r_a^i}{|\Br_a|^{d}}\frac{r_b^j}{|\Br_b|^{d}}=
\frac{A_{d-1}}{2(d-2)|\Bx_a-\Bx_b|^{d-2}}
\left[\delta^{ij}-(d-2)
\frac{(x_a^i-x_b^i)(x_a^j-x_b^j)}{|\Bx_a-\Bx_b|^2}\right]
\end{equation}

\begin{eqnarray}
I_{ab}&\equiv&\partial_{a}^i\partial_{b}^i\int
d^d\Bx\sum_c\sum_d~\frac{1}{|\Br_c|^{d-2}}\frac{1}{|\Br_d|^{d-2}} \nn
&=&2(d-2)A_{d-1}\left[-\delta_{ab}\sum_c\frac{1}{|\Bx_a-\Bx_c|^{d-2}}+
\frac{1}{|\Bx_a-\Bx_b|^{d-2}}\right]
\end{eqnarray}

\begin{equation}
\sum_a I_{aa}=
-2(d-2)A_{d-1}\mathop{\sum\sum}_{a\Neq b}\frac{1}{|\Bx_a-\Bx_b|^{d-2}}
\end{equation}

\begin{eqnarray}
& &\sum_a\sum_b I_{ab}I_{ab} \nn
&=&4(d-2)^2A_{d-1}^2\left[
\mathop{\sum\sum\sum}_{b\Neq a, c\Neq
a}\frac{1}{|\Bx_a-\Bx_b|^{d-2}}\frac{1}{|\Bx_a-\Bx_c|^{d-2}}+
\mathop{\sum\sum}_{a\Neq b}\frac{1}{|\Bx_a-\Bx_b|^{2(d-2)}}\right]
\end{eqnarray}

\begin{eqnarray}
& &\partial_{a}^i\partial_{b}^i\int
d^d\Bx\sum_c\sum_d\sum_e~\frac{1}{|\Br_c|^{d-2}}\frac{1}{|\Br_d|^{d-2}}
\frac{1}{|\Br_e|^{d-2}} \nn
&=&3(d-2)A_{d-1}\left[-\delta_{ab}\sum_c\sum_d\frac{1}{|\Bx_a-\Bx_c|^{d-2}}
\frac{1}{|\Bx_a-\Bx_d|^{d-2}}\right. \nn
& &\qquad\qquad\qquad
+\frac{1}{|\Bx_a-\Bx_b|^{d-2}}\sum_c\left(\frac{1}{|\Bx_a-\Bx_c|^{d-2}}+
\frac{1}{|\Bx_b-\Bx_c|^{d-2}}\right) \nn
& &\qquad\qquad\qquad
\left.-\sum_c\frac{1}{|\Bx_a-\Bx_c|^{d-2}}\frac{1}{|\Bx_b-\Bx_c|^{d-2}}
\right]
\end{eqnarray}

\begin{eqnarray}
& &\sum_a\partial_{a}^i\partial_{a}^i\int
d^d\Bx\sum_c\sum_d\sum_e~\frac{1}{|\Br_c|^{d-2}}\frac{1}{|\Br_d|^{d-2}}
\frac{1}{|\Br_e|^{d-2}} \nn
&=&-3(d-2)A_{d-1}\left[
\mathop{\sum\sum\sum}_{b\Neq a, c\Neq
a}\frac{1}{|\Bx_a-\Bx_b|^{d-2}}\frac{1}{|\Bx_a-\Bx_c|^{d-2}}+
\mathop{\sum\sum}_{a\Neq b}\frac{1}{|\Bx_a-\Bx_b|^{2(d-2)}}\right]
\end{eqnarray}


\end{document}